\documentclass[12pt,preprint]{aastex}

\def\spose#1{\hbox to 0pt{#1\hss}}

\def\etal{{\em et al.}}

\def\arcsec{\hbox{$^{\prime\prime}$}}

\def\lsim{\mathrel{\hbox{\rlap{\lower.55ex \hbox {$\sim$}}\kern-.0em\raise.4ex \hbox{$<$}}}} 
\def\gsim{\mathrel{\hbox{\rlap{\lower.55ex \hbox {$\sim$}}\kern-.0em\raise.4ex \hbox{$>$}}}}

\def\grb{GRB\thinspace{980329}}
\def\ts{\thinspace}

\begin{document}

\title{The Broadband Afterglow of GRB\thinspace{980329}}

\author{S. A. Yost\altaffilmark{1},
 D. A. Frail\altaffilmark{1,2},
 F. A. Harrison\altaffilmark{1},
 R.    Sari\altaffilmark{3},
 D.    Reichart\altaffilmark{1},
 J. S. Bloom\altaffilmark{1},
 S. R. Kulkarni\altaffilmark{1},
 G. H. Moriarty-Schieven\altaffilmark{4}
 S. G. Djorgovski\altaffilmark{1},
 P. A. Price\altaffilmark{1,5},
 R. W. Goodrich\altaffilmark{6},
 J. E. Larkin\altaffilmark{7},
 F.    Walter\altaffilmark{1}
 D. S. Shepherd\altaffilmark{2}, 
 D. W. Fox\altaffilmark{1}
 G. B. Taylor\altaffilmark{2}, 
 E.    Berger\altaffilmark{1},
 T. J. Galama\altaffilmark{1}
}

\begin{abstract}

We present radio observations of the afterglow of the bright
$\gamma$--ray burst \grb\ made between one month and several years
after the burst, a re-analysis of previously published submillimeter
data, and late-time optical and near-infrared (NIR) observations of
the host galaxy. From the absence of a spectral break in the
optical/NIR colors of the host galaxy, we exclude the earlier
suggestion that \grb\ lies at a redshift of $z\gsim5$.  We combine our
data with the numerous multi-wavelength observations of the early
afterglow, fit a comprehensive afterglow model to the entire broadband
dataset, and derive fundamental physical parameters of the blast-wave
and its host environment. Models for which the ejecta expand
isotropically require both a high circumburst density and extreme
radiative losses from the shock. No low density model ($n \ll 10$
cm$^{-3}$) fits the data. A burst with a total energy of $\sim
10^{51}$ erg, with the ejecta narrowly collimated to an opening angle
of a few degrees, driven into a surrounding medium with density $\sim
20$ cm$^{-3}$, provides a satisfactory fit to the lightcurves over a
range of redshifts.

\end{abstract}

\slugcomment{Submitted to The Astrophysical Journal}

\altaffiltext{1}{Division of Physics, Mathematics and Astronomy,
  105-24, California Institute of Technology, Pasadena, CA 91125}

\altaffiltext{2}{National Radio Astronomy Observatory, P.O. BOX `O', Socorro,
NM 87801} 

\altaffiltext{3}{Theoretical Astrophysics 130-33, California Institute
  of Technology, Pasadena, CA 91125}

\altaffiltext{4}{National Research Council of Canada, Joint Astronomy Centre, 
660 N. A'ohoku Place Hilo, HI 96720}

\altaffiltext{5}{Research School of Astronomy and Astrophysics, Mount 
Stromlo Observatory, via Cotter Rd., Watson Creek 2611, Australia}

\altaffiltext{6}{W. M. Keck Observatory, 65-1120 Mamalahoa Highway,
Kamuela, HI 96743}

\altaffiltext{7}{Department of Physics and Astronomy, University of
  California, Los Angeles, CA 90095}


\section{Introduction}

The discovery of long-lived afterglow emission from gamma-ray bursts
(GRBs) has revolutionized the field by enabling redshift
determinations and host galaxy identifications.  In addition, by
interpreting the emission in the theoretical framework of a
relativistic shock wave expanding into a circumburst medium, broadband
afterglow measurements can constrain the explosion geometry and
energetics, as well as the properties of the surrounding medium
\citep{wg99, gps99c, hys+01, pk01}. The
hydrodynamic evolution of the blast-wave is strongly influenced by the
the total energy in the shock, the geometry of the outflow and the
density structure of the medium into which it is expanding.  The
time-dependence of the radiated emission from the shock depends on
the hydrodynamic evolution, as well as on the partition of energy
between the radiating electrons and the magnetic field. With data of
sufficient quality in conjunction with a theoretical model, broadband
measurements of the spectral evolution of the afterglow allow us to
deduce fundamental physical parameters of the explosion.

Bright gamma-ray bursts such as \grb\ have been targets of extensive
broadband followup.  \grb\ was well-localized in the gamma-rays, and
the position quickly refined as a result of the X-ray detection by
In't Zand {\em et al.}  (1998)\nocite{iaa+98}. However, initial
searches for an optical afterglow were unsuccessful until variable
emission was identified at radio wavelengths
\citep{tfk+98}. Subsequent observations of the radio position
uncovered a faint optical counterpart \citep{gcn41}, as well as a
relatively bright near-infrared (NIR) transient \citep{gcn43}.
Because of the delay in the identification of the optical afterglow,
the early optical monitoring was somewhat sparse.  In spite of the
eventual detection of optical and NIR counterparts, no redshift has
been determined, due to the faintness of the host emission and its
lack of prominent emission lines.

In this paper we present broadband observations of the afterglow of
\grb.  We monitored the emission at multiple radio frequencies over
times extending beyond the first month (early time data were reported
by Taylor \etal\ (1998)). In addition, we present late-time optical
and NIR observations in the R, I, H and K bands, as well as a
re-analysis of submillimeter data reported by Smith
\etal\ (1999)\nocite{stv+99}. We fit the broadband emission to a fireball
model, allowing us to derive the physical parameters of the
afterglow, and measure the properties of the host galaxy.

\section{Observations and Data Reduction}\label{sec:obs}

\subsection{Optical/NIR Data}

Optical and NIR observations were made using the
Keck\footnotemark\footnotetext{The W.~M.~Keck Observatory is operated
by the California Association for Research in Astronomy, a scientific
partnership among California Institute of Technology, the University
of California and the National Aeronautics and Space Administration.}
10-m telescopes on Mauna Kea, Hawaii in the R, I, and K bands. Except
for a re-calibration of the K-band points from Larkin {\em et al.}
(1998a,b)\nocite{lgd+98a}\nocite{lgd+98}, all the photometric
measurements presented in Table~\ref{tab:optical} have not been
previously published.  Deep R- and I-band images were obtained on
several epochs using the Low Resolution Imaging Spectrometer (LRIS;
\citet{occ+95}), and on one epoch (2001 January) with the Echelle
Spectrograph and Imager (ESI; \citet{smbs00}). The data consist of
multiple CCD exposures with typical durations of 300~s taken in good
photometric conditions at an airmass between 1.1 to 1.3.  We flat
fielded and combined the data following standard practice.  For
calibration purposes, we used an image of the \grb\ field taken at the
Palomar 60-inch in January 1999 under photometric conditions. Its
photometric zeropoint was determined using four field stars which were
in common with Reichart \etal~(1999)\nocite{rlm+99}.

The K-band images of the \grb\ field were taken with the Near Infrared
Camera (NIRC; Matthews \& Soifer 1994\nocite{ms94}) on the Keck I
telescope.  We used IRAF to reduce the data, and the DIMSUM
package\footnotemark\footnotetext{http://iraf.noao.edu/iraf/ftp/contrib/dimsumV2/}
to combine images and subtract sky background.  We used observations
of Persson \etal\ (1998)\nocite{pmk+98} standards on each photometric
night to calibrate a sequence of stars in the field, against which we
performed relative photometry of the afterglow.  We estimate that the
calibration is accurate to approximately 5\%.

We also made use of an H-band observation from the Hubble Space Telescope
({\it HST}) archive. HST observed \grb\ on 1998 October 16 through the
F160W filter with NICMOS Camera 2 as part of GO7863 (PI: A. Fruchter).  We
used the standard NICMOS pipeline developed at STScI with the best
available reference and calibration files to process the images.  We
then shifted and added the exposures to form a final, stacked image.
We used the IRAF aperture photometry tools to measure the
afterglow flux and the NICMOS photometric calibration published by
Stephens \etal\ (2000)\nocite{sfo+00} to convert this measurement to
the standard H-band magnitude.

\subsection{Very Large Array (VLA)} 

Observations were made with the VLA\footnotemark\footnotetext{The NRAO
is a facility of the National Science Foundation operated under
cooperative agreement by Associated Universities, Inc.  NRAO operates
the VLA.} at three frequencies, 8.46 GHz, 4.86 GHz and 1.43 GHz. All
observations employed the standard VLA continuum mode, with data being
recorded in two adjacent 50 MHz bandpasses. Calibration of the array
phase was derived from observations of the nearby calibrators
J0653+370 and J0713+438.  Calibration of the flux density scale was
done using J1331+305 or J0542+498, or by extrapolating past
measurements of J0713+438, which was has a very stable flux density.
Table \ref{tab:Table-VLA} contains a log of the observations and a
list of the measured flux densities.  The VLA data for the first month
after the burst were published by Taylor et al.~(1998\nocite{tfk+98}).

\subsection{Owens Valley Radio Observatory (OVRO)} 

A continuum observation with OVRO was made at a central frequency of
99.98 GHz on 1998 Dec 13.42 and 14.40 UT, several months after the
burst.  The observations totalled approximately 14 hours on source in
good 3\,mm weather conditions. The total bandwidth was 2 GHz,
resulting in an rms of $\sim0.7$~mJy. The flux was calibrated using
the extragalactic source 3C273, and the phase from the nearby quasar
0552+398.

A further observation was made at a central frequency of 99.5 GHz on
2001 November 26.48 UT, nearly 1400 days after the burst.  The
observation consisted of a single, 11-hr long track (9 hours on
source) taken under excellent 3\,mm conditions with four antennas. The
total bandwidth was 4 GHz, resulting in an rms of $\sim0.5$~mJy. We
set the flux density scale using the extragalactic sources 3C84 as
well as 3C273, and derived phase calibration from J0646+448.

\subsection{James Clark Maxwell Telescope (JCMT)}

We re-analyzed archival JCMT\footnotemark\footnotetext{The JCMT is
operated by The Joint Astronomy Centre on behalf of the Particle
Physics and Astronomy Research Council of the UK, the Netherlands
Organization for Scientific Research, and the National Research
Council of Canada.} observations of this burst taken at frequencies of
650 GHz, 350 GHz and 220 GHz. Details of the observing procedure can
be found in Smith {\em et al.} (1998)\nocite{stv+99}. The data were
reduced using the SCUBA User Reduction Facility \citep{jl98} in the
same manner as that for the recent GRB\,010222 \citep{fbm+01}.  Raw
signals were flat-fielded to account for the small differences in
bolometer response, extinction corrected, and de-spiked to remove
anomalous signals above the 3-sigma level.  Short time-scale sky
variations were also removed using pixels around the edge of the array
containing no source emission \citep{jlh98}.  A flux calibration
factor was then applied to convert to Jy. Flux calibration factors
(FCF) of 240 $\pm$ 15 Jy/V, 197 $\pm$ 13 Jy/V and 384 $\pm$ 82 Jy/V were
applied to the 220 GHz, 350 GHz and 660 GHz data, respectively
\citep{cou00}. Table \ref{tab:Table-JCMT} contains a log of the
observations and a list of the measured flux densities.

\section{Afterglow Model}\label{sec:fitting}

We interpret the data in the framework of the cosmological fireball
model ({\em e.g.}, \cite{pir99}), in which an energetic
explosion accelerates a small amount of matter to ultrarelativistic
velocities. Internal shocks within this flow produce the burst event
itself, while the relativistic shock propagating into the surrounding
medium produces the afterglow.

In the model, we assume the ultrarelativistic shock transfers
a constant fraction of its total energy to the magnetic field
($\epsilon_{B}$) as well as a constant fraction to shocked,
thermalized electrons ($\epsilon_{e}$). As in strong
subrelativistic shocks, the electrons are assumed to be accelerated
into a power--law distribution of energies (P$(\gamma_{e}) \propto
\gamma^{-p}$) and they radiate via synchrotron emission under the
influence of magnetic fields. This produces a broken power--law
spectrum whose peak and spectral breaks evolve in time according to
the shock's behaviour \citep{spn98}, set by the total energy in the
shock, the geometry of the ejecta and the density of matter
surrounding it.

The specific model we used to fit the data is outlined in some detail
in Harrison \etal\ (2001)\nocite{hys+01}. It includes the effects of
inverse Compton (IC) scattering on the shock evolution and emitted
radiation spectrum as prescribed by Sari \& Esin
(2001)\nocite{se01}. It allows for a conical (jet-like) outflow
geometry with half-opening angle $\theta_{jet}$ (using the treatment
given by Sari, Piran \& Halpern (1999)\nocite{sph99}), and for
expansion into a medium of constant density $n$, or a medium with a
density gradient, {\em i.e.}, $n\propto r^{-2}$. The latter density
profile would be typical of a medium altered by the wind of a massive
star (in this case the GRB progenitor).  In addition, we calculate and
include radiative corrections to the shock energy. This is a refined
method as compared to the adiabatic evolution typically assumed in our
previous work \citep{hys+01, bdf+01, bsf+00}.  We
approximate the radiative evolution by treating the shock at each
moment as though it were instantaneously adiabatic, with an energy
calculated at that particular time. This is an appropriate treatment
as long as losses are moderate, with the change in energy being slow.

In addition to the basic input physics describing the evolution of the
fireball outlined above, the broadband model for GRB afterglows
incorporates several additional effects resulting from propagation of
the radiation between the fireball and the observer. These include
interstellar scintillation in the centimeter radio regime, dust
extinction in the optical/NIR regime, and a contribution to the
emission from the host galaxy of the GRB.  These features are evident
in the data, and must be included to derive accurate model parameters.

Interstellar scintillation (ISS), due to the turbulent ionized gas of
our Galaxy distorting wavefronts propagating to the observer, can be
important at radio wavelengths \citep{goo97, fkn+97}. We
account for ISS by first estimating the fractional variation in the
flux density expected by these distortions \citep{wal97}.  This
uncertainty in the model flux is added in quadrature to the
statistical uncertainties in the measured values when estimating
$\chi^2$.

Dust within GRB host galaxies (either in the circumburst environment
or along the line of sight) is a source of extinction for optical/NIR
afterglows \citep{ksm+00, sfc+01, dfk+01} which must be
accounted for in modeling the optical spectrum. The frequency
dependence of the extinction curve is uncertain, since we know little
about the ISM properties of the \grb\ host galaxy. For \grb , the
observed red optical to near-IR afterglow spectrum \citep{ppm+98}
suggests a steep extinction curve. The best characterized (relatively)
steep extinction law is for the Small Magellanic Cloud (SMC). We
therefore adopt the SMC bar extinction curve from Weingartner \&
Draine (2001)\nocite{wd01} , which is parameterized by the extinction
level $A_{V}$ in the rest frame of the host galaxy. Other extinction
curves were used in model attempts and did not substantially change
the results.
 
Depending on its luminosity, the host galaxy of the GRB may contribute
a background level which dominates the total brightness at late
times. This is observed as a flattening of the light curves, chiefly
in the optical and IR.  For \grb , we include a constant term for the
R, I, H and K to represent the host emission, and the values of these
terms are fit by the model.  The J-band only includes a single data
point, so we interpolate the host term between the I and H
fits. Evidence for host emission has also been observed at centimeter
wavelengths \citep{bkf01}, again deduced from late-time flattening of
the lightcurve.  The radiation at these wavelengths would most likely
result from synchrotron and thermal bremsstrahlung in galaxies
undergoing substantial star formation \citep{con92}.  We include the
possibility of centimeter host emission in our model, where we fit for
the normalization at 1.43 GHz, and scale other bands as
$\nu^{-0.8}$. In light of recent claimed detections of GRB hosts at
submillimeter wavelengths \citep{hlm+00, bkf01, fbm+01, bgcn1182+01,
berger+02}, we consider the possibility of a submillimeter host, and
allow for such a component in the fit, with a modified blackbody as in
Frail \etal\ (2001)\nocite{fbm+01}, parameterized by its flux at 350
GHz. This last host component was found not to be required for our
best model.

\section{Best-Fit Broadband Model}\label{sec:ag}

The afterglow model described in \S\ref{sec:fitting} was fit to the
data summarized in Tables \ref{tab:optical}-\ref{tab:Table-JCMT}. In
addition, we included all previously published data in the X-ray
\citep{iaa+98}, optical \citep{ppm+98, gcp+99, rlm+99}, and radio
\citep{tfk+98} bands. We converted the X-ray measurements to flux
values using a photon index of 2.4. We corrected the optical data for
absorption in our Galaxy \citep{sfd98} before converting to flux
densities using the factors in Bessell (1979) \nocite{b79} for the
optical and Bessell \& Brett (1988)\nocite{bb88} for the near-IR
bands. To account for any cross-calibration uncertainty, we have added
in quadrature 5\% uncertainties to all the measured fluxes.

The model which best describes this broadband dataset is a collimated
outflow expanding into a constant density medium.  The best model
parameters, derived from least-squares maximizing the fit probability,
are summarized in Table~\ref{tab:model} and
Figs.~\ref{fig1}-\ref{fig6}. The $\chi^2$ for the fit is 116.4 for 92
degrees of freedom. Although we derived fits for three representative
redshifts ($z = 1, 2, 3$), the results of which are all shown in
Table~\ref{tab:model}, we confine our detailed comments below to the
$z=2$ solution. This choice of representative redshifts was made based
upon the range of likely $z$ for this burst. Very high redshifts
$z~\gsim~5$ are not considered as they are not compatible with the
underlying host's colors, as detailed in \S\ref{sec:host}. A redshift
$z<1$ is considered implausible due to the lack of lines expected to
be detected (if $z<1$) in several spectra taken of the host. The host
is visible at optical wavelengths, and thus not completely obscured,
so it is quite unlikely that the prominent star-formation-related
oxygen line [OII]$\lambda$3727 or the H$\alpha$ line would not have been
observable if the host is a faint galaxy at $z<1$. Many
of the basic conclusions do not depend on the redshift, or can be
easily scaled given the information below. In
\S\ref{sec:alter} we discuss some of the limitations of our best fit,
as well as some alternate models which also fit the data,
but only with unphysical parameters.

In the best model for $z=2$, the isotropic-equivalent fireball
energy at the time when the fireball evolution becomes nearly
adiabatic ($E_{\rm iso}(t_{\nu_c = \nu_m})$) is approximately
$10^{54}$~ergs.  The measured gamma-ray fluence for this GRB is
$F_\gamma=5.5\times 10^{-5}$ erg cm$^{-2}$ (\cite{iaa+98}), so the
isotropic gamma-ray energy is $E_{\rm iso}(\gamma)= 4\pi F_\gamma
d_L^2(1+z)^{-1}\simeq 6\times{10}^{53}$ erg. The large energy budgets
inferred for both the shock $E_{\rm iso}(t_{\nu_c = \nu_m})$ and the
emitted gamma-ray radiation $E_{\rm iso}(\gamma)$ derived assuming
isotropy are greatly reduced in this model by the relatively large
degree of collimation (jet opening angle of $\theta_{jet}\sim
2^\circ$). A similar degree of collimation has been inferred previously in
GRB afterglows: GRB\ts{990510} has $\theta_{jet}=3^\circ$
\citep{hbf+99} and GRB\ts{000911} has $\theta_{jet}=2^\circ$
\citep{pbk+02}. For a two-sided jet this implies a total energy in
the fireball shock of $8.3 \times 10^{50}$~ergs, similar to the energy
released in supernovae. Likewise, for $z=2$ the geometry-corrected
gamma-ray energy is reduced to $4.0\times 10^{50}$ erg, a value that
is in good agreement with the mean of $5\times 10^{50}$ erg derived
from a larger sample \citep{fks+01}.  We note the total energy is
similar ($5.4 \times 10^{50}$) for $z=1$, and a factor of about three
higher for $z=3$.

The energy quoted in Table \ref{tab:model} is a lower limit on the
true initial energy of the blastwave since it is derived at a time
$t_{\rm \nu_{c} = \nu_{m}}$=6.1 d, when the lowest energy electrons
can cool within the dynamical timescale of the system.
Observationally, this corresponds to the time when the cooling break
$\nu_{c}$ crosses the spectral break which results from the low-energy
cutoff in the input electron energy spectrum, $\nu_{m}$.  This
criterion separates the two regimes of radiative losses; early times
when radiant energy results in a decrease in the blastwave energy with
time, and late times when the blast-wave evolution is adiabatic.  For
our best model, radiative losses are important, but not extreme
even at early times, since the fraction of energy in radiating
electrons is not dominant ($\epsilon_{e}$=12\%).  We estimate that
from the time the GRB ends ($\sim$~10-100s post-trigger), to when the
blastwave is nearly adiabatic, the energy drops by a factor of five.
If we restrict the interval to begin when the first afterglow data
were measured, ($t=0.25$ d for the first data point to $t_{\rm \nu_{c}
= \nu_{m}}$), the energy drops by a factor of 1.6.  From the time
$t=t_{\rm \nu_{c} = \nu_{m}}$ until late times the energy drops by
only 15\% .

The energy in the fireball derived from our model significantly
exceeds the emitted gamma-ray energy, {\em i.e.,} $E_{\rm iso}(t_{\nu_c = \nu_m})
>E_{\rm iso}(\gamma)$.  As discussed above, the
initial fireball energy will be even larger if radiative losses are
taken into account.  This is the case for the majority of GRB
afterglows with energies derived from model fits (see {\em e.g.}
\cite{pk01b}). In the fireball model, it is likely that the energy 
remaining in the shock during the afterglow exceeds that of the prompt
gamma-ray emission of the GRB event itself; this will be the case for
a radiative efficiency of $<$ 50\% during the GRB. The radiative
efficiency of internal shocks driving the prompt GRB are expected to
be $\sim$~10\%, not $\gg$~50\%, leaving most of the initial shock
energy in the fireball \citep{gsw01}.

The ratio of the energy fraction in magnetic field,
$\epsilon_{B}=$17\%, to that in the electrons, $\epsilon_{e}$=12\%,
determines the relative importance of Comptonization.  Compton
scattering can contribute significantly to the total cooling rate if
the ratio of electron to magnetic energy fractions is greater than
unity \citep{se01}.  In our best model for \grb , this ratio is
of order unity, so Comptonization does not dominate the electron
cooling, but it is a non-negligible effect.  Flux from Compton
scattering can in fact be seen peaking in the X-rays in
Figs.~\ref{fig4}~and~\ref{fig6}, largely as a result of the steep electron
spectral index.  For an electron energy spectral index of 2, with
equal energies in each logarithmic frequency interval (the
infinite-energy limit), the IC luminosity would be lower than the
synchrotron luminosity at all frequencies, and no IC peak would be
observable in the spectrum. The index of $p=2.9$ derived for
the best model puts less energy in each successive decade above
the peak frequency. With a significant circumburst electron density
providing a non-negligeable opacity to Compton scattering, the peak of
the IC flux density, above the synchrotron peak, dominates the total
flux near the X-rays.

The circumburst medium density derived from the model, $n=20$
cm$^{-3}$, is comparable to that of GRB\ts{000926} \citep{hys+01} and
that of several other bursts (\cite{pk01b}). This relatively high
density, in reference to an average galactic ISM, can be inferred from
the measured value of the self-absorption break $\nu_{a}$
\citep{gps99c}, which is shown in Fig.\ref{fig6}.  The frequency
$\nu_a$ depends upon other fundamental parameters besides the ambient
density ($\nu_a\propto n^{3/5} \epsilon_{e}^{-1}
\epsilon_{B}^{1/5} E_{iso}^{1/5}$), and in this particular case 
the high $\nu_a$ results from the relatively high density, combined
with a moderate electron energy fraction.  Models with low circumburst
densities ({\em i.e.}, $n \ll 10$ cm$^{-3}$) cannot be fit to the data
by adjusting the energy and electron fraction. Highly radiative
models, with large electron energy fractions, can provide reasonable
fits to the data, but the densities (depending on the redshift of the
fit) vary from approximately the same as to an order of magnitude
greater than those for the best model (see \S\ref{sec:alter}).

\subsection{Host Galaxy}\label{sec:host}

We infer the presence of measurable flux from the GRB host galaxy in
the optical from the late-time flattening of the lightcurves
(Fig.~\ref{fig3}).  In the final epoch of our optical observations the
measured brightness of R=26.53$\pm$0.22 (see Table \ref{tab:optical})
is essentially entirely due to the host, and is only 1.5 mag fainter
than the median R magnitude for known GRB host galaxies
\citep{dfk+01b}. Our measured magnitude differs from the preliminary
value of R=27.7$\pm$0.3 given by Holland {\em et al.}
(2000)\nocite{hth+00}.  This discrepancy seems to be due to a
mis-identification of the host in their HST images which they identify
with a source 0.5\arcsec\ southwest of the GRB position. Further work
by that group \citep{jah02} has identified the host at a position
consistent with that of Bloom {\em et al.}  (2002)\nocite{bkd01},
which established with improved astrometry the host whose R-band
magnitude is given here. Bloom {\em et al.}'s (2002)\nocite{bkd01}
measured offset for the GRB from the host center is only 37$\pm$48
mas, corresponding to a host-normalized offset of 0.215$\pm$0.291,
placing \grb\ within the half light radius of its compact host.  Our I
band host flux values from Table \ref{tab:optical} are in good
agreement with Jaunsen {\em et al.}'s (2002)\nocite{jah02} late-time I
band host measurement.

In Fig. \ref{fig5}\ we plot the spectral energy distribution of the
afterglow at 0.7 days after the burst along with the late-time
measurements from Table \ref{tab:optical}, assumed to be due to the
host galaxy. We corrected all points for extinction in our Galaxy.
Palazzi {\em et al.} (1998)\nocite{ppm+98} first noted the steep
spectral slope between the R and I bands seen in the early time
afterglow, and Fruchter (1999)\nocite{fru99} suggested that this
``dropout'' of the R band could be produced by absorption from the
Ly$\alpha$ forest if the redshift $z$ of this burst was greater than
five. This steep slope, however, is not reflected in the host
spectrum.  The afterglow at 0.7 days after the burst is significantly
redder (R$-$I=2.7$\pm$0.4) than the host galaxy (R$-$I=0.2$\pm$0.3)
itself (note that the quoted R$-$I above are corrected for Galactic
extinction).  From the absence of a strong $R-I$ break in the host
spectrum, we can rule out a redshift of $z\gsim{5}$ for \grb .

Another result that emerges from the modeling is the presence of
significant dust extinction in the host galaxy. Early attempts to
model the optical data for this burst \citep{ppm+98, lcr99} also found
that the spectrum was substantially reddened by dust.  Our fitted host
$A_{\rm V}$ corresponds to hydrogen column density of $N_{H}\simeq
2\times 10^{21}$ cm$^{-2}$, assuming a gas-to-dust ratio similar to
that of the Milky Way \citep{ps95, rei01}.  In't Zand {\em et al.}
(1998)\nocite{iaa+98} used the X-ray spectrum of the afterglow to
derive a column density $N_{H}=1.0\pm0.4\times10^{22}$ cm$^{-2}$, with
a 99\% confidence range of $1.3\times10^{21}$-$1.5\times10^{22}$
cm$^{-2}$, after subtracting a Galactic contribution of approximately
0.8$\times$10$^{21}$cm$^{-2}$ \citep{dl90}.  To translate this measured column
into the restframe of the host galaxy requires multiplying by a factor
$(1 + z)^{8/3}$.  For a redshift of 2, this exceeds the value derived
from optical extinction. Such discrepancies have been noted before
\citep{vgo+99, gw01}, and are taken as evidence of significant dust
destruction in the circumburst medium out to a radius of order 10-20
pc \citep{wd00, fkr01, rei01b}. In the case of \grb , however, the
redshift is not known, and the uncertainties in the dust extinction
law make it difficult to claim evidence for dust destruction.

There is a suggestion of a 1.43 GHz radio host in the data, with the
model requiring a flat, positive component on average.  Although the
addition of this component improves the fit, the significance of a
nonzero radio host parameter is only 3$\sigma$.  If real, a radio host
at $\simeq 25~\mu$Jy would be about 1/3 the host flux density found
for GRB\ts{980703} \citep{bkf01}.  We note that the submillimeter
data are in good agreement with the afterglow model  (see Fig.\ref{fig2}),
and we do not require any host contribution in this band.

\section{Alternate Models and Limitations}\label{sec:alter}

The best fit model given in \S\ref{sec:ag}, while not a unique
interpretation of the data, is a self-consistent solution that derives
reasonable values for the blastwave energy $E_{\rm
iso}(t_{\nu_c=\nu_m})$, the opening angle $\theta_{jet}$, the ambient
density $n$ and the parameters of the shock ($p$, $\epsilon_e$,
$\epsilon_B$). A collimated outflow expanding into a constant density
medium describes {\it all} the data well, addressing puzzling features
described in previous attempts to fit this afterglow
\citep{fru99,lcr99,stv+99} without invoking a very high redshift or
other additional components.  The unusual features of this data set
include the submillimeter excess, the very red $R-I$ afterglow colors
and their relation to the host, and the observed decline of the peak
flux density $F_m$ with time (or equivalently with decreasing
frequency $\nu_m$, since $\nu_m$ decreases with time).

\begin{figure}
	\epsscale{0.85}
  \plotone{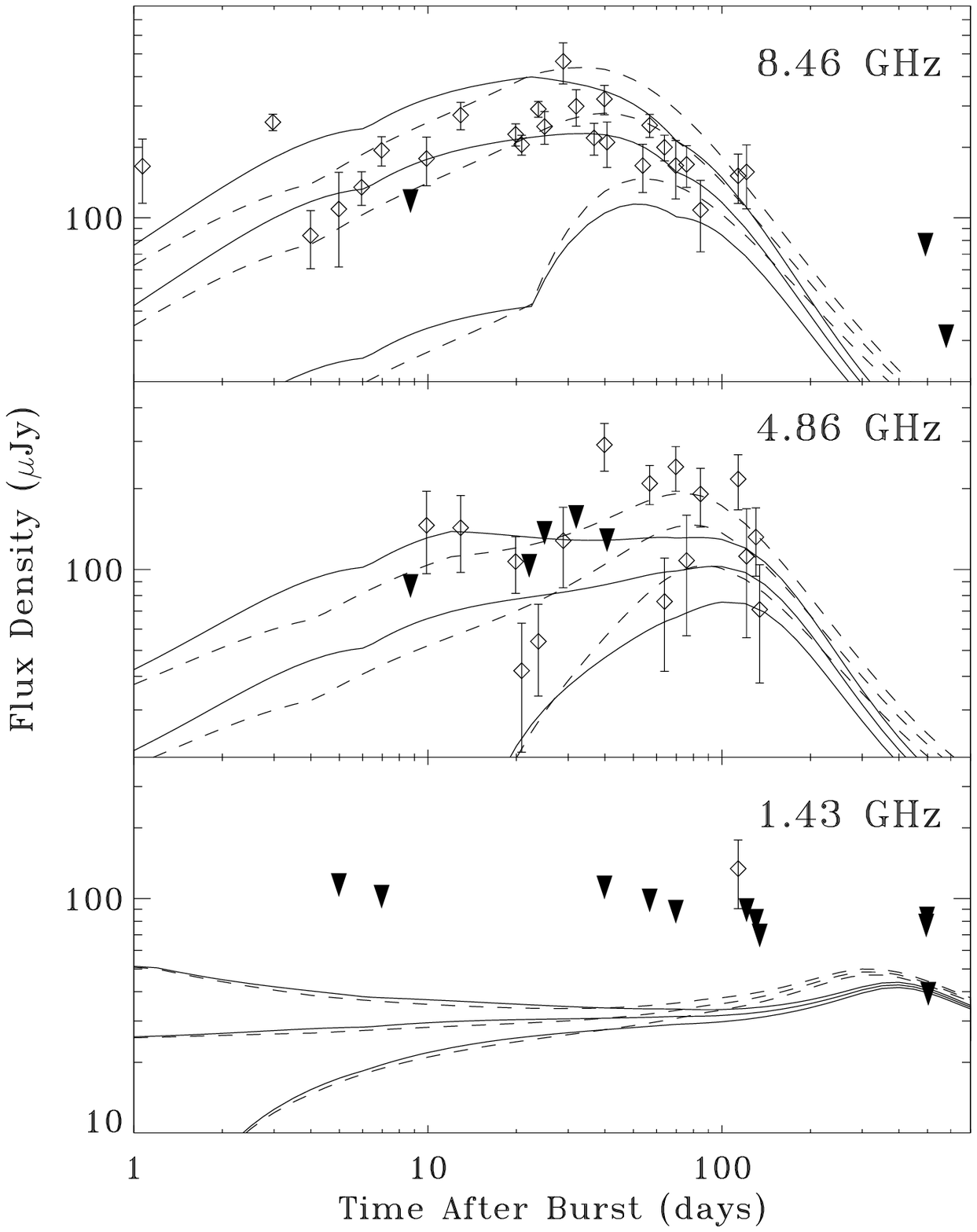}
\caption{Radio lightcurves of the GRB 980329 afterglow. Both the best
  model and the extreme radiative solution, described in
  \S\ref{sec:alter} are plotted. The light curves of the ``best''
  model (the best physical model; see \S\ref{sec:ag} and \ref{sec:alter}
  for details) are solid; the radiative solution's are dashed. The
  model lightcurves are plotted with their calculated 1-$\sigma$
  scintillation envelopes above and below. Data that are not at least
  detected at the 2-$\sigma$ level are presented as 2-$\sigma$ upper
  limits (max(flux density, 0)~+~2~$\times$~rms noise; black
  triangles). The 1.43 GHz data is only significant as a whole. Note
  that the 8.46 GHz data at $\le$ 3 days, which was not included in
  the fits, is significantly in excess of both models.
\label{fig1}}
\end{figure}

\begin{figure}
\epsscale{1.}
\plotone{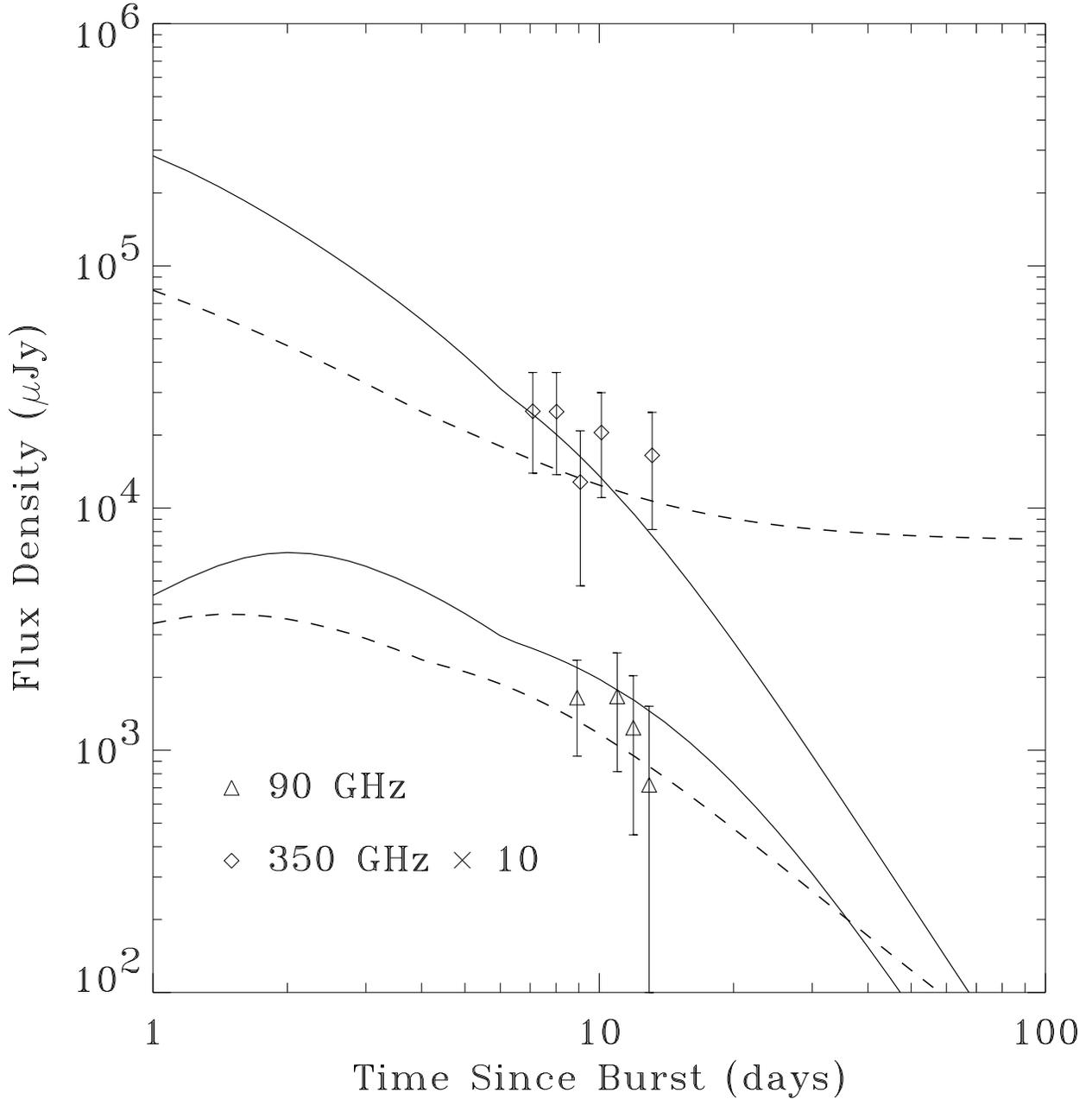}
\caption{Millimeter and submillimeter lightcurves of the GRB 980329
  afterglow; the 350 GHz data and model are multiplied by 10 for
  clarity. The ``best'' model (the best physical model; see
  \S\ref{sec:ag} and \ref{sec:alter} for details) is shown with solid
  lightcurves; the radiative solution (\S\ref{sec:alter}) with dashed
  ones. The best model fits the reanalyzed data without the need to
  include a submillimeter host component. The radiative solution is
  plotted with the submillimeter host component, required to account
  for $\sim$ 1/2 of the 350 GHz flux. See the text for model details.
\label{fig2}}
\end{figure}

This last feature warrants further explanation, since it is an effect
that can only result from a finite number of physical causes.  The
``peak flux cascade'' can be readily seen in Figs. \ref{fig1} and
\ref{fig2}, where the peak flux density is 2.5 mJy at 350 GHz but
declines to 1.5 mJy at 90 GHz and further falls to 0.35, 0.2 and
$<$0.1 mJy at 8.46 GHz, 4.86 GHz, and 1.43 GHz, respectively. A fit to
the data near maximum between 4.86 GHz and 350 GHz gives a power-law
slope of 0.59$\pm$0.07.  A similar behavior was also observed for the
afterglow of GRB\ts{970508} \citep{gwb+98b, fwk00}. Within the
context of the standard fireball model there are three ways to produce
this behavior. First, if the flux evolution is observed after
collimation of the ejecta becomes evident in the light curve decay
({\em i.e.}, post-jet), then $F_{m} \propto \nu_{m}^{1/2}$
(\cite{sph99}). For our best model, this is what produces the
observed peak flux cascade.  Alternately, a density gradient $n
\propto r^{-2}$ in the surrounding medium (a stellar wind model) will
give $F_{m} \propto \nu_{m}^{1/3}$ \citep{cl99}.  Finally, radiative
losses can produce a peak flux cascade. However, unless these are severe
(namely most of the shock energy in electrons), the effect is quite
weak.  For example, for $\epsilon_e\simeq 0.1$, $F_{m} \propto
\nu_{m}^{0.08}$, while for $\epsilon_e\simeq 1$, $F_{m} \propto
\nu_{m}^{0.37}$ \citep{cps98}.

\begin{figure}
\plotone{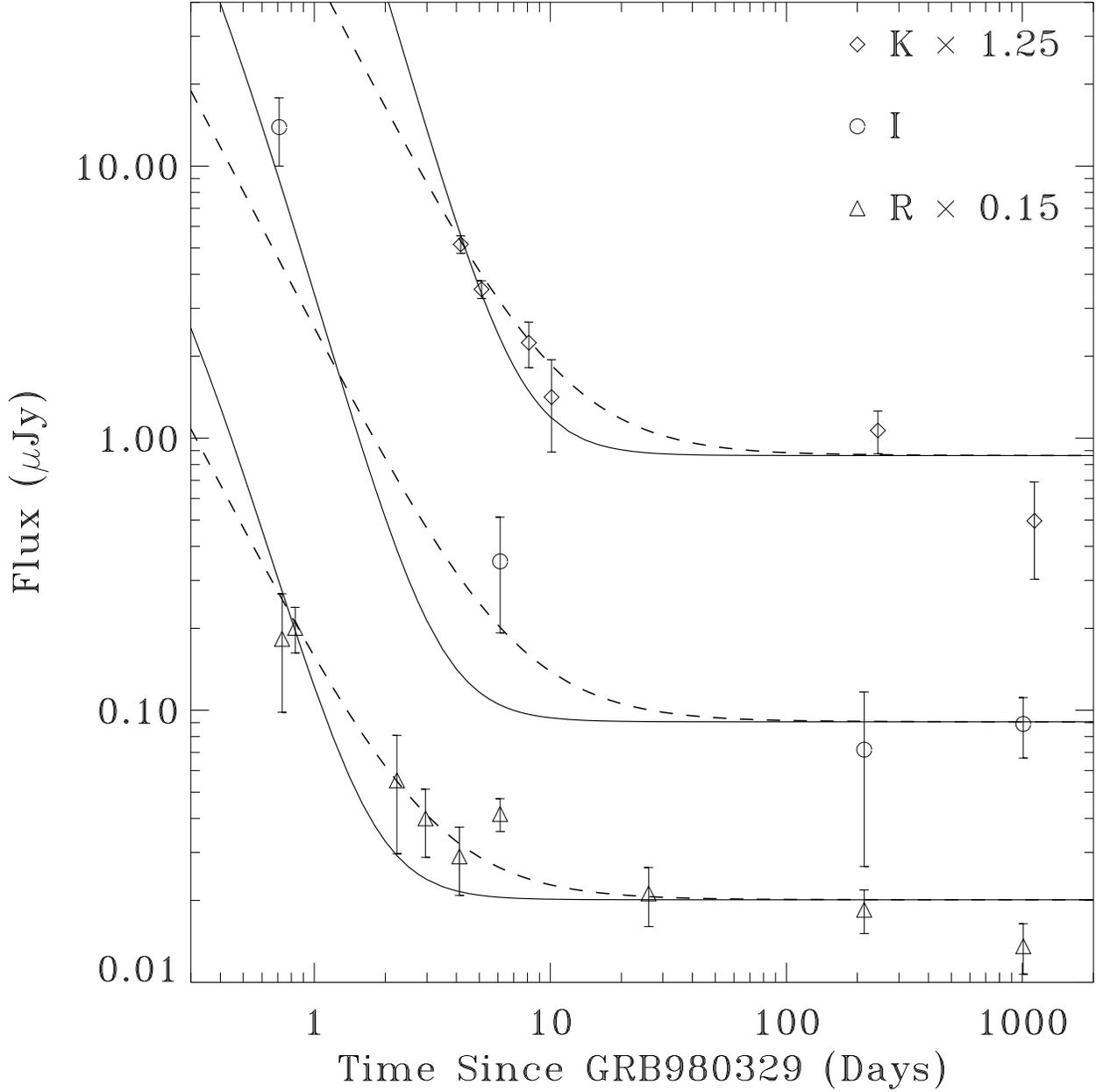}
\caption{Optical lightcurves of the GRB 980329 afterglow at R, I and K
  bands. The ``best'' model (the best physical model; see
  \S\ref{sec:ag} and \ref{sec:alter} for details) is shown with solid
  lightcurves; the extreme radiative solution (\S\ref{sec:alter}) with
  dashed ones. The data are corrected for Galactic (but not host)
  extinction. The late-time host fluxes can be clearly seen.
\label{fig3}}
\end{figure}

\begin{figure}
\plotone{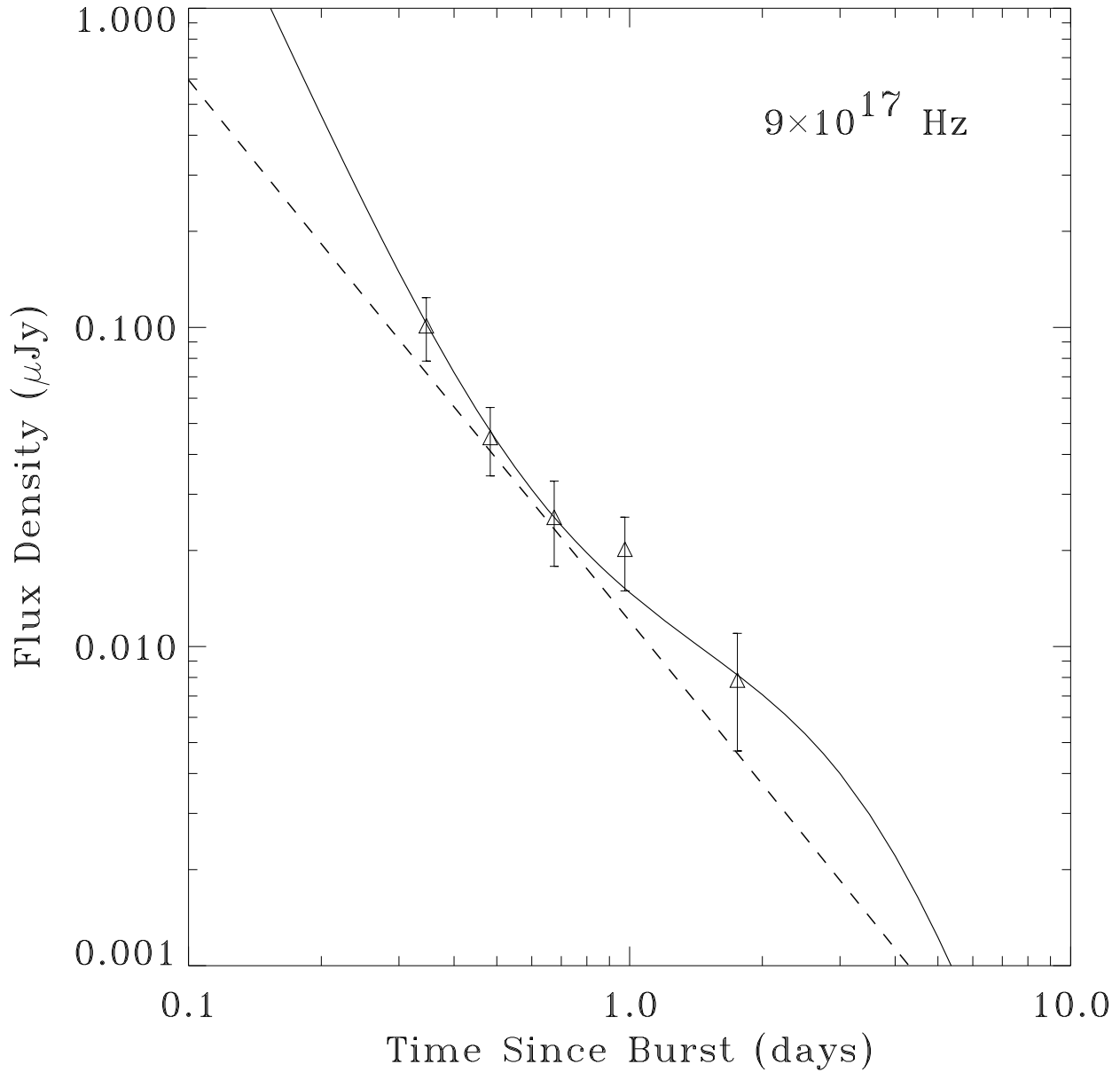}
\caption{The X-ray lightcurve of the GRB 980329 afterglow, with both
  models.  The ``best'' model (the best physical model; see
  \S\ref{sec:ag} and \ref{sec:alter} for details) is shown with solid
  lightcurves; the extreme radiative solution (\S\ref{sec:alter}) with
  dashed ones (see text for details). The curvature seen in the best
  model after 2 days is the signature of a significant inverse Compton
  contribution to the X-ray afterglow flux at that time.\label{fig4}}
\end{figure}

\begin{figure}
\plotone{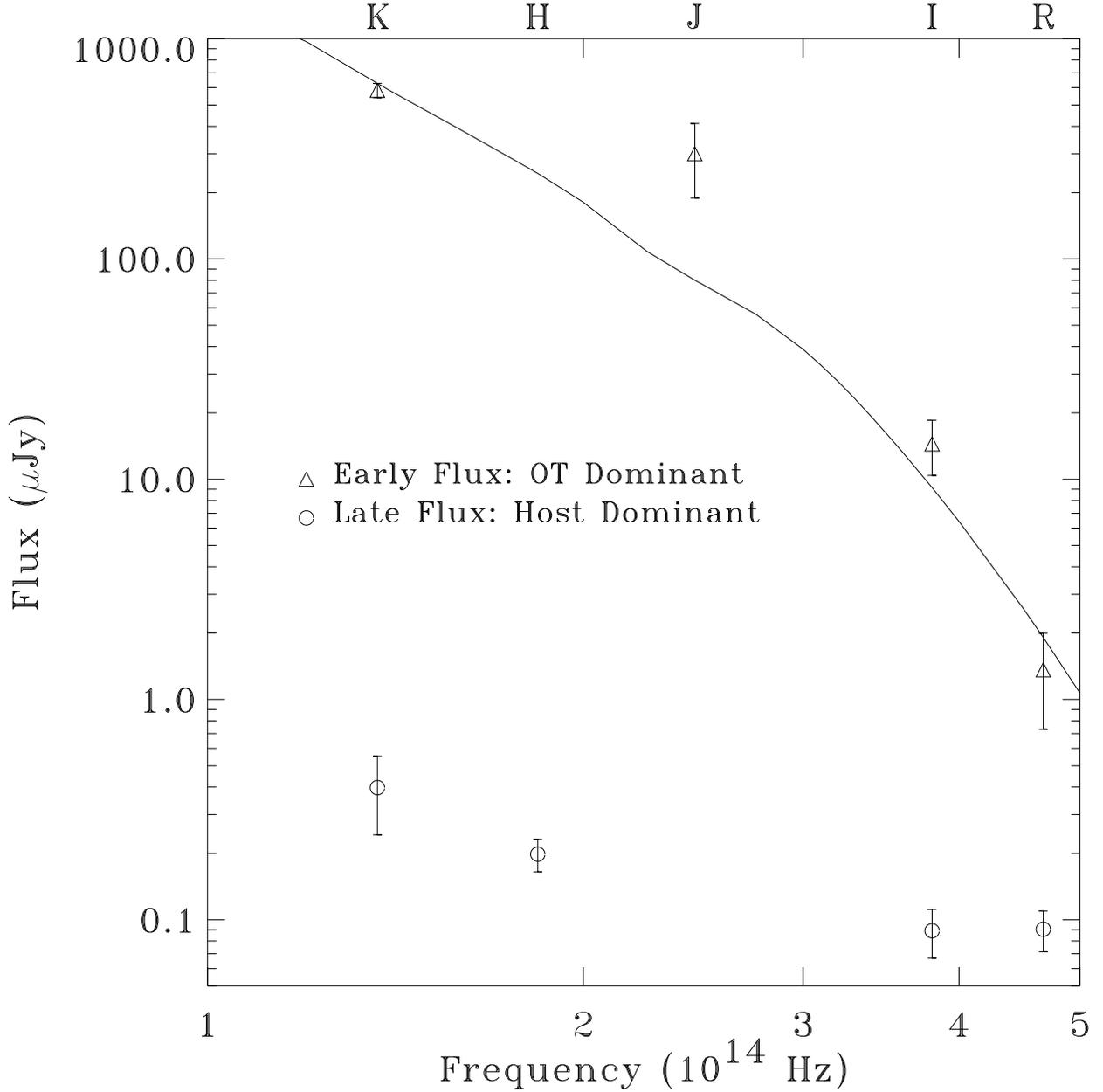}
\caption{Comparison of afterglow and underlying host optical flux
  densities.  The data are corrected for Galactic (but not host)
  extinction.  The first available data points (open triangles) in R,
  I, J, and K bands (at 0.73, 0.71, 8.1 and 4.2 days post-burst
  respectively) were each scaled to 0.7 days using our best afterglow
  model, which is overplotted.  The afterglow flux dominates these
  points and the spectral steepness from I to R is clearly seen.
  (Note that the J band point, being extrapolated from a time when
  the host flux was beginning to become important, is a less reliable
  afterglow flux indicator.) The late time measurements (open circles)
  at R, I, H, and K bands are also plotted to show the host
  spectrum. The host spectrum does not show the steep spectral slope
  between the I and R bands, as expected if \grb\ was at $z \gsim 5$
  (see text for details).
\label{fig5}}
\end{figure}

\begin{figure}
\plotone{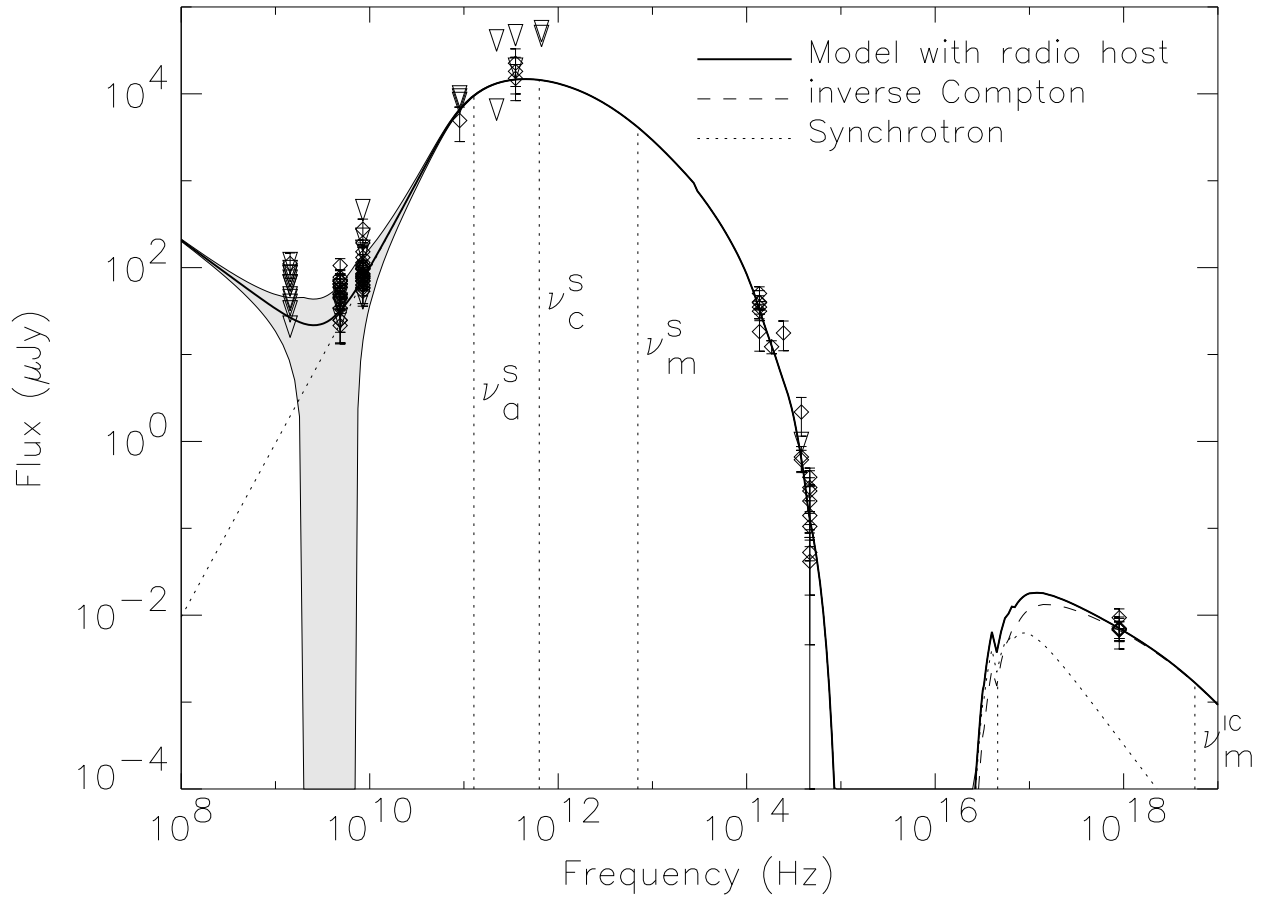}
\caption{All of the data, scaled to day 2 post-burst, and plotted on
our model's day 2 spectrum. The spectrum's inverse Compton and
synchrotron flux components are decomposed from the total. The grey
shaded region represents the estimated flux uncertainty in the model
of the observed spectrum at this time, due to interstellar
scintillation. The high self-absorption frequency and predominance of
the Comptonized flux at X-ray frequencies can be clearly
seen. \label{fig6}}
\end{figure}

The other models we derived that fit the primary characteristics of
the data all required unusual physical assumptions.  For example, we
found a solution with extreme radiative corrections (100\% of the
shock energy going into electrons) that could reproduce the observed
peak flux cascade. Formally, this model fits the data better than the
``best model'' presented in the previous section, however, it reaches
the unphysical edge of parameter space, and with extreme radiative
losses our near-adiabatic treatment breaks down and cannot be fully
trusted. The highly radiative model is isotropic, and for $z=2$, has
the following parameters: $E_{\rm iso}(t_{\nu_c=\nu_m}) \simeq 2\times
10^{52}$~ergs, $n \simeq 20$~cm$^{-3}$, $p\simeq 2.02$,
$\epsilon_B\simeq 0.17$, $\epsilon_e \rightarrow 1$, A(V) $\simeq$ 1.1
and a centimeter host of $\simeq$ 17 $\mu$Jy at 1.43 GHz. In addition
to the unphysical assumption about the electron energy partition, this
model also has an electron spectral index approaching two, and hence a
diverging total energy.  This radiative model only accounts for
$\approx$ 1/2 of the 350 GHz flux, suggesting an underlying
submillimeter host of $\simeq$ 0.7 mJy (this component improves the
fit at $\approx$~3~$\sigma$ level). This submillimeter host flux level
is just below the sensitivity limit of current instruments and would
likely not be detectable at late times, if it indeed exists. We
consider the collimated solution presented in \S\ref{sec:ag} to be the
best model as it is the best fit of the models with realistic
parameters. This best model reproduces the flux cascade with a
relatively narrow collimation angle (early jet-break).  We note that
the $p = 2.88$ we derive for the best $z = 2$ model is somewhat larger
than found for other afterglows, which generally fall in the range $p
= 2.2 - 2.4$.  It is, however, physically reasonable, and we regard
all the parameters associated with the best model as acceptable. The
highly radiative solution is plotted along with the best model in the
lightcurves presented in Figs.~\ref{fig1}--\ref{fig4}. The fit is
visibly somewhat better, but at the cost of unphysical assumptions
concerning the underlying parameters of the fireball.

The most serious limitation to modeling the afterglow of \grb\ is the
lack of a good redshift estimate. Even a fairly comprehensive dataset
such as this cannot constrain fundamental parameters without knowing
the distance.  This is because the synchrotron emission can be
reproduced at different $z$ simply by rescaling the physical
parameters by appropriate powers of $(1+z)$. Only ``second order''
effects such as host extinction, the IC component and radiative
corrections do not directly re-scale with $(1 + z)$.  In principle, we
could include $z$ as a free parameter in the model, and fit for the
best value. In practice, however, the combination of the sparseness of
the real dataset, and the uncertainties in the model prevent any
unique redshift determination. This is evident from
Table~\ref{tab:model}, where we show a good fit with reasonable
physical parameters for all three redshifts.

The absence of a strong break in the host galaxy spectral colors
(Fig.~\ref{fig5}) allows us to place an upper limit on the host
redshift. Spectral energy distribution (SED) fittings to host colors
may place a stronger constraint ({\em e.g.}~\cite{castlamb99},
concerning the 970228 event), especially if the factor of $\sim$ four
decline from K-band to I-band can be modelled by a Balmer
break. Further HST imaging in optical bands might prove fruitful in
this regard.

The absence of a redshift is responsible in part for the relatively
large and uncertain estimate of the electron index $p$ in our best
model.  The post-jet evolution of the optical light curves is
determined by $p$ \citep{sph99}. Unfortunately, this data is sparsely
sampled at early times (Fig.~\ref{fig3}), prior to when the host
galaxy dominates the light.  Likewise, the index $p$ determines the
shape of the synchrotron spectrum, and should therefore be derivable
from the measurements.  At optical wavelengths, however, there is a
degeneracy between $p$ and the dust extinction law --- the latter of
which can depend sensitively on $z$. Lacking any knowledge about the
extinction properties of dust in high redshift galaxies we adopted the
SMC extinction law.  Neither the X-ray spectral slope, nor the X-ray
to optical flux ratio can break this degeneracy, since the
contribution from inverse Compton scattering alters the X-ray flux
normalization as well as the spectrum.

Finally, we note that both our best solution and the highly radiative
solution fail to predict the early radio emission at 8.46 GHz ($t<3$
d). This level of fluctuation is too great to be accounted for by the
estimated ISS effects.  Prompt, short-lived radio emission in excess
of the normal afterglow component has been detected toward other GRBs
\citep{kfs+99, fbg+00, hys+01}. This is usually attributed to
radiation from a reverse shock \citep{sp99b}, and we suggest that
this may explain the bright early-time radio point for \grb .

\section{Conclusions}

The new observations presented here, along with the data in the
literature, form a complete collection of broadband measurements for the
afterglow from the bright burst \grb\ beginning from early to late times.
We have used these data together with a comprehensive fireball model to
deduce fundamental physical parameters of the event (energy, geometry,
density) and to measure the properties of the host galaxy.

The late-time optical/NIR data shows that the host is significantly
bluer than the afterglow. Thus we reject the hypothesis that the very
red afterglow colors were due to Lyman-$\alpha$ absorption in the
intergalactic medium to a very high redshift source.
 
All of the afterglow's features can be explained over a wide range of
$z$ by a model in which the ejecta are collimated in a jet.
Significant dust extinction is inferred within the host galaxy and a
moderately high circumburst density $n \simeq 20$ cm$^{-3}$ is
required.  Although this collimated model is not a unique solution to
the data, it explains the red optical/NIR color and the cascade in the
peak flux from submillimeter to centimeter wavelengths (see
\S\ref{sec:alter}) without resorting to an extreme redshift $z \gsim
5$ or requiring additional complications in the host galaxy's
properties.  Models which invoke isotropic geometry require such
complications, circumburst densities up to 10 times higher, as well
as large, unphysical radiative corrections.  Correcting the
isotropic-equivalent gamma-ray energy release for the collimation, for
$z = 2$ we obtain $E(\gamma) = E_{\rm iso}(\gamma)\theta^{2}/2 = 4.0
\times 10^{50}$~ergs.  This value is typical of other events to date
and is easily accounted for by current progenitor models.

\acknowledgements 

RS acknowledges support from the Fairchild Foundation and from a NASA
ATP grant. JSB acknowledges a grant from the Hertz foundation.
Research with the Owens Valley Radio Telescope, operated by Caltech,
is supported by NSF grant AST96-13717


\begin{deluxetable}{lcccc}
\footnotesize \tablecaption{Optical/NIR Observations of \grb
\label{tab:optical}}
\tablewidth{0in} 
\tablehead{
\colhead{Epoch} &
\colhead{ } & 
\colhead{ } &
\colhead{ } &
\colhead{ } \\
\colhead{(UT)} &
\colhead{Telescope} &
\colhead{Inst.} &
\colhead{Band} & 
\colhead{Mag.}
}
\tablecolumns{3} 
\startdata
1998 Apr. 04.29 & Keck & LRIS & R & 25.31$\pm$0.14 \\  
1998 Apr. 24.25 & Keck & LRIS & R & 26.04$^{+0.30}_{-0.23}$ \\  
1998 Nov. 29    & Keck & LRIS & R & 26.19$\pm$0.19 \\  
2001 Jan. 01.32 & Keck & ESI  & R & 26.53$^{+0.25}_{-0.20}$ \\
\hline
1998 Apr. 04.29 & Keck & LRIS & I & 24.79$^{+0.65}_{-0.40}$ \\
1998 Nov. 29    & Keck & LRIS & I & 26.52$^{+1.07}_{-0.53}$ \\
2001 Jan. 01.32 & Keck & ESI  & I & 26.28$^{+0.31}_{-0.24}$ \\
\hline
1998 Oct. 16 & HST & NICMOS & H & 24.32$^{+0.19}_{-0.16}$ \\
\hline
1998 Apr 02.33  & Keck & NIRC & K & 20.50$\pm$0.06 \\
1998 Apr 03.27  & Keck & NIRC & K & 20.91$\pm$0.06 \\
1998 Nov 28.49  & Keck & NIRC & K & 22.21$\pm$0.19 \\
2001 Apr 27.25  & Keck & NIRC & K & 23.04$^{+0.53}_{-0.36}$ \\
\enddata
\end{deluxetable}

\begin{deluxetable}{lcc}
\footnotesize
\tablecolumns{3}
\tablewidth{0pc}
\tablecaption{VLA Observations of \grb\ \label{tab:Table-VLA}}
\tablehead {
\colhead {Epoch} &
\colhead{Frequency}&
\colhead{Flux Density}\\
\colhead {(UT)} &
\colhead{(GHz)}&
\colhead{($\mu$Jy)}}
\startdata
1998 May 04.92  & 8.46   & 219$\pm$32 \\
1998 May 08.01  & 8.46   & 321$\pm$43 \\
1998 May 08.84  & 8.46   & 210$\pm$45 \\ 
1998 May 21.97  & 8.46   & 167$\pm$38 \\ 
1998 May 24.96  & 8.46   & 248$\pm$25 \\
1998 May 31.98  & 8.46   & 200$\pm$23 \\
1998 Jun. 06.84 & 8.46   & 167$\pm$46 \\
1998 Jun. 12.98 & 8.46   & 169$\pm$33 \\
1998 Jun. 21.77 & 8.46   & 108$\pm$36 \\
1998 Jul. 20.70 & 8.46   & 151$\pm$35 \\
1998 Jul. 28.43 & 8.46   & 157$\pm$47 \\
1999 Aug. 03.60 & 8.46   &  40$\pm$23 \\
1999 Oct. 29.43 & 8.46   &  13$\pm$11 \\
2000 Sep. 01.40 & 8.46   &$-8.0\pm$14 \\
\hline
1998 May 08.01  & 4.86   & 291$\pm$57 \\
1998 May 08.84  & 4.86   &  60$\pm$41 \\
1998 May 24.96  & 4.86   & 209$\pm$33 \\
1998 May 31.98  & 4.86   &  76$\pm$34 \\
1998 Jun. 06.84 & 4.86   & 241$\pm$44 \\
1998 Jun. 12.98 & 4.86   & 108$\pm$51 \\
1998 Jun. 21.77 & 4.86   & 191$\pm$46 \\
1998 Jul. 20.70 & 4.86   & 217$\pm$49 \\
1998 Jul. 28.43 & 4.86   & 112$\pm$56 \\
1998 Aug. 06.68 & 4.86   & 132$\pm$37 \\
1998 Aug. 10.58 & 4.86   &  71$\pm$33 \\
\hline
1998 May 08.01  & 1.43   &  47$\pm$39 \\
1998 May 24.96  & 1.43   &  42$\pm$34 \\
1998 Jun. 06.84 & 1.43   & 0.5$\pm$49 \\
1998 Jul. 20.70 & 1.43   & 134$\pm$43 \\
1998 Jul. 28.43 & 1.43   &  16$\pm$42 \\
1998 Aug. 06.68 & 1.43   & $-39\pm$45 \\
1998 Aug. 10.58 & 1.43   & $-17\pm$39 \\
1999 Aug. 06.68 & 1.43   &  36$\pm$25 \\
1999 Aug. 09.63 & 1.43   &  40$\pm$26 \\
1999 Aug. 14.65 & 1.43   &$-1.7\pm$22 \\
\enddata                                                                 
\end{deluxetable}                

\begin{deluxetable}{lcc}
\footnotesize
\tablecolumns{3}
\tablewidth{0pc}
\tablecaption{Millimeter Observations of \grb\ \label{tab:Table-JCMT}}
\tablehead {
\colhead {Epoch } &
\colhead{Frequency}&
\colhead{Flux Density\tablenotemark{a}} \\
\colhead {(UT)} &
\colhead{(GHz)}&
\colhead{(mJy)}}
\startdata
1998 Apr. 5.25  &  650 & $-9.9 \pm 9.3$  \\
1998 Apr. 6.18  &  650 & $-11.8 \pm 8.9$ \\
1998 Apr. 7.22  &  650 & $0.1 \pm 7.5 $ \\
1998 Apr. 8.27  &  650 & $0.4 \pm 8.5$  \\
1998 Apr. 11.30 &  650 & $-1.1 \pm 5.1$ \\
\hline
1998 Apr. 5.25  &  350 & $2.51 \pm 1.11$ \\   
1998 Apr. 6.18  &  350 & $2.50 \pm 1.12$ \\
1998 Apr. 7.22  &  350 & $1.28 \pm 0.80$ \\
1998 Apr. 8.27  &  350 & $2.05 \pm 0.94$  \\
1998 Apr. 11.30 &  350 & $1.65 \pm 0.83$ \\ 
\hline
1998 Apr. 7.30  &  220 & $3.69 \pm 2.18$ \\
1998 Apr. 8.16  &  220 & $-0.96 \pm 1.04$ \\ 
\hline
1998 Dec. 13.91 & 100.0 & $0.04 \pm 0.70$ \\  
\hline
2001 Nov. 26.48 & 99.5 & $0.54 \pm 0.50$ \\  
\tablenotetext{a}{Due to our recalibration of the data, the submillimeter 
fluxes presented here are in disagreement with Smith \etal\ 
(1999)'s\nocite{stv+99} values from the same observations, and in good 
agreement with the afterglow model with no host excess (see Fig.\ref{fig2}). }
\enddata
\end{deluxetable}

\begin{deluxetable}{cccc}
\footnotesize
\tablecolumns{4}
\tablewidth{0pc}
\tablecaption{Fit parameters for assumed $z=1, 2, 3 $ \label{tab:model}}
\tablehead{
\colhead{Parameter} & 
\colhead{z=1} & 
\colhead{z=2} &
\colhead{z=3}}
\startdata
$\chi^2$ for 105 data pts & 113.1 & 116.4 & 119.4\\
$t_{\rm jet}$ (days) & 0.21 & 0.12 & 0.29\\
$t_{\rm non rel.}$ (days) & 35 & 70 & 96\\
$t_{\rm \nu_{c} = \nu_{m}}$ (days) & 2.4 & 6.1 & 10.0\\
$E_{\rm iso}(t_{\nu_c = \nu_m})(10^{52}$~erg)\tablenotemark{a} & 15 & 126
& 107\\
$n (cm^{-3})$ & 20 & 20 & 29\\
$p$  & 2.55 & 2.88 & 3.06\\
$\epsilon_e$ (fraction of E) & 0.08 & 0.12 & 0.14\\
$\epsilon_B$ (fraction of E) & 0.27 & 0.17 & 0.08\\
$\theta_{jet} (rad)$ & 0.081 & 0.036 & 0.049\\
\hline
host A(V) & 2.8 & 1.9 & 1.4\\
host R ($\mu Jy$) & 0.13 & 0.13 & 0.13\\
host I ($\mu Jy$) & 0.090 & 0.091 & 0.090\\
host H ($\mu Jy$) & 0.20 & 0.20 & 0.20\\
host K ($\mu Jy$) & 0.68 & 0.69 & 0.70\\
host 1.4 GHz ($\mu Jy$) & 19 & 25 & 26\\
\hline 
$E_{\rm iso}(\gamma)$\tablenotemark{b} (10$^{53}$~erg) & 1.5$\pm$0.2 & 5.1$\pm$0.6 & 9.5$\pm$1.1\\
$E(\gamma)$ (10$^{50}$~erg)\tablenotemark{c} & 5.0$\pm$0.6 & 3.3$\pm$0.4 & 11$\pm$1\\ 
\enddata
\tablenotetext{a}{Isotropic equivalent blastwave energy (not corrected for
collimation)}
\tablenotetext{b}{Isotropic-equivalent energy emitted in
the gamma-rays by the GRB, if it occurred at this redshift, calculated by 
the method of Bloom et al. (2001)\nocite{bfs01}}
\tablenotetext{c}{The isotropic-equivalent energies given above, 
corrected assuming the jet angles (without uncertainty) presented above}
\end{deluxetable}

\end{document}